\documentclass{article}
\usepackage{amsmath,amsthm}
\usepackage{amssymb,latexsym}
\usepackage[mathscr]{eucal}
\usepackage{setspace}
\usepackage{graphics}
\usepackage{array}

\newcommand{\pa}{\partial}
\newcommand{\pr}{\prime}
\newcommand{\al}{\alpha}
\newcommand{\na}{\nabla}
\newcommand{\eps}{\epsilon}
\newcommand{\lp}{\left(}
\newcommand{\rp}{\right)}
\newcommand{\lb}{\left[}
\newcommand{\rb}{\right]}
\newcommand{\lc}{\left\{}
\newcommand{\rc}{\right\}}
\newcommand{\la}{\left<}
\newcommand{\ra}{\right>}
\newcommand{\be}{\begin{equation}}
\newcommand{\ee}{\end{equation}}
\newcommand{\bs}{\boldsymbol}

\newcommand{\ihat}{\bf\hat{i}}

\begin{document}

\title{\textbf{Turbulent Relative Particle Dispersion}}         
\author{B. K. Shivamoggi\\
University of Central Florida\\
Orlando, FL 32816-1364, USA\\
}        
\date{}          
\maketitle

\noindent \large{\bf Abstract} \\ \\
In this paper, phenomenological developments are used to explore several aspects of the relative particle dispersion (RPD) in different physical fully-developed turbulence (FDT) situations. The role played by the FDT cascade physics underlying this process is investigated. Many of these aspects are motivated by previous laboratory experiment and numerical simulation results. These are,
\begin{itemize}
  \item[*] spatial intermittency effects exhibiting,
  \begin{itemize}
    \item[(a)] reduction of RPD in 3D FDT, corroborating the numerical simulation results (Boffetta and Sokolov \cite{Bof});
    \item[(b)] prevalence of power-law scaling of RPD in 2D FDT enstrophy cascade (no matter how weak spatial intermittency effects are), corroborating the difficulty in observing Lin \cite{Lin} exponentical scaling law in laboratory experiments (Jullien \cite{Jul});
  \end{itemize}
  \item[*] quasi-geostrophic FDT aspects exhibiting an enhanced RPD in the baroclinic regime of the energy cascade and a negative eddy-viscosity development to shed some insight into this aspect;
  \item[*] quasi-geostrophic FDT aspects exhibiting particle clumping in the baroclinic regime of the enstrophy cascade;
  \item[*] reduction of RPD, development of the ballistic regime  and particle clustering due to compressibility effects in FDT, corroborating the laboratory experiment and numerical simulation results (Cressman et al. \cite{Cre}).
\end{itemize}

\noindent These results are developed from the established scaling relations for the various physical FDT cases and are further validated via alternative dimensional/scaling developments for the various physical FDT cases similar to the one given for 3D FDT by Batchelor and Townsend \cite{Bat1}.

\pagebreak

\noindent\Large\textbf{1. Introduction}\\

\large The concept of turbulent diffusion by the continuous movement of a single particle was introduced by Taylor \cite{Tay}, who defined a diffusion coefficient $D$ such that the mean square displacement $\la \lb s (t) \rb^2 \ra$ is given by
\be\tag{1}
\la \lb s (t) \rb^2 \ra = 2Dt
\ee
$t$ being the elapsed time\footnote{$t$ is assumed to be large enough so the conditions of the initial release have been forgotten.}. The single-particle diffusion characterizes the translational motion of a small cloud released at $t=0$ in a turbulent flow, while the rate of spreading of this cloud is measured by the average rate at which two particles moving with this cloud separate due to turbulent advection. Large-scale translating motions cause the two particles to meander together while small-scale straining motions cause them to drift apart. Richardson \cite{Ric}, therefore, proposed that turbulent diffusion should instead be characterized by the distance between neighboring particles, and hence by the effective shear that acts over this distance rather than the magnitude of the turbulent velocity field\footnote{Indeed, as Taylor \cite{Tay2} mentioned, ``Richardson was a very interesting and original character who seldom thought on the same lines as his contemporaries and often was not understood by them."}. If the interparticle distance is much greater than the integral scale $L$, the particles are uncorrelated and hence separate diffusively; the particles separate chaotically if the interparticle distance is much less than the Kolmogorov microscale $\eta$. On the other hand, if the interparticle distance is within the inertial range (where there is no characteristic length scale), the properties of turbulent flows are characterized only by the mean energy dissipation rate $\varepsilon$. Consequently, one may then expect to find {\it universal} super-diffusive\footnote{The super-diffusive behavior is caused by an increasing range of eddy sizes that contribute to the relative velocity of the two particles as the interparticle-separation increases with $t$ (see equation (3)).} behavior in the relative particle dispersion (RPD) process, which may be interpreted in terms of an interparticle separation dependent turbulent relative diffusivity.  From a purely empirical analysis of atmospheric dispersion data, Richardson \cite{Ric} then surmised that the turbulent relative diffusivity defined by the rate of increase of the mean square interparticle separation distance $R\lp t\rp$,

\be\tag{2}
\mathscr{D} \equiv \frac{1}{2} ~\frac{d}{dt} \la \lb R (t) \rb^2 \ra
\ee

\noindent goes like 4/3 power of this distance\footnote{On theoretical grounds, (3) can be justified only if the advecting velocity field is finite correlated in time. This follows by noting, from the initial-value problem,

\be\notag
\left.
\begin{matrix}
(d/dt) R (t) = V (t)\\
t = 0 ~: ~R = R_0
\end{matrix}
\rc
\ee

\noindent that

\be\notag
\frac{1}{2} \la \lb R (t) - R_0 \rb^2 \ra = \int\limits_0^t \int\limits_0^t \la V \lp t^\pr \rp V \lp t^{\pr \pr} \rp \ra dt^\pr dt^{\pr \pr}.
\ee

\noindent Assuming a stationary process, this gives

\be\notag
\la \lb R (t) - R_0 \rb^2 \ra = 2t \int\limits_0^t C (\tau) d \tau
\ee

\noindent where $C (\tau)$ is the velocity autocorrelation,

\be\notag
C (\tau) \equiv \la V (0) V (\tau) \ra.
\ee

\noindent If $C (\tau)$ decays in time fast enough so $\int\limits_0^\infty C (\tau) d \tau$ is finite, we have a diffusive motion, with turbulent relative diffusivity,

\be\notag
\mathscr{D} = \lim\limits_{t \Rightarrow \infty} \frac{1}{2t} \la \lb R (t) - R_0 \rb^2 \ra = \int\limits_0^\infty C (\tau) d \tau.
\ee

\noindent On the other hand, if $C (\tau)$ decays slowly in time, we have a super-diffusive motion (Bouchaud and Georges \cite{Bou}) with

\be\notag
\la \lb R (t) - R_0 \rb^2 \ra \sim t^{2 \nu}, ~\nu > 1/2
\ee

\noindent and

\be\notag
\mathscr{D} \sim \la \lb R (t) - R_0 \rb^2 \ra^{\frac{2 \nu - 1}{2 \nu}}.
\ee},

\be\tag{3}
\mathscr{D} \sim \la \lb R (t) \rb^2 \ra^{2/3}.
\ee

\noindent implying that the turbulent relative dispersion is an accelerating process because as the particles separate further, the relative dispersive-motion scales (bounded above by $R\lp t\rp$) become larger. The most important application of two-particle dispersion is in understanding the motion of passive scalars (like pollutants) in atmospheres and oceans.

Obukhov \cite{Obu} showed that Richardson's relation (3) can be derived via Kolmogorov's \cite{Kol} (K41) theory\footnote{The K41 theory stipulates that the probability distribution function of velocity fluctuations, in the inertial range, depends only on the mean energy dissipation rate $\varepsilon$ and the length scale $\ell$. Further, $\varepsilon$ is independent of the Reynolds number $R_e$.} for homogeneous isotropic 3D fully developed turbulence (FDT). When the interparticle separation is within the inertial range, Obukhov \cite{Obu} gave\footnote{One might think there is some logical inconsistency in the expression in (4) - $\mathscr{D}$ and $\la \lb R (t) \rb^2 \ra$ are Lagrangian quantities while $\varepsilon$ is defined fully precisely in the Eulerian formulation. However, there exists a fully Lagrangian way to define $\varepsilon$ - it is (-4) times the Lagrangian time derivative of the squared relative velocity between the two fluid particles (Falkovich et al. \cite{Fal}).}

\be\tag{4}
\mathscr{D} \sim \varepsilon^{1/3} \la \lb R (t) \rb^2 \ra^{2/3}
\ee

\noindent hence the name Richardson-Obukhov (RO) scaling\footnote{Taylor \cite{Tay2}, as implied by (4), actually suggested that Richardson ``had the idea that the index (4/3) was determined by something connected with the way energy was handed down from larger to smaller and smaller eddies".}. Implicit in (4) is the assumption that the effective shear acting across $R$ in 3D FDT arises from eddies of size $\sim R$ (Kraichnan and Montgomery \cite{Kmont}). It may be mentioned that Richardson \cite{Ric} indicated that $\mathscr{D}$ is proportional to $\varepsilon^{1/3}$ and that $\mathscr{D}$, like $\varepsilon$, remains nearly independent of the Reynolds member $R_e$. So, Richardson \cite{Ric}, interestingly provided the first experimental evidence for {\it dissipative anomaly} (i.e., $\varepsilon$ is almost independent of $R_e$) via the irreversibility of turbulent RPD (Falkovich and Frishman \cite{FalFrish}) and came very close to fully discovering the basic tenets of the K41 theory 15 years earlier. On the other hand, the Richardson formulation connects with the universal aspects of FDT actually stronger than the K41 formulation since the unphysical effects due to sweeping by large scales L are precluded from the outset in the Richardson formulation\footnote{Indeed, as Monin and Yaglom \cite{Mon} remarked, Richardson's formulation ``indicates Richardson's belief in the existence of a universal physical law of sufficiently simple form."}.

In this paper, phenomenological developments are used to explore several aspects of RPD in different physical FDT situations. The role played by the FDT cascade physics underlying this process is investigated. Many of these aspects are motivated by previous laboratory experiment and numerical simulation results. These are,

\begin{itemize}
  \item[*] spatial intermittency effects exhibiting,
  \begin{itemize}
    \item[(a)] reduction of RPD in 3D FDT, corroborating the numerical simulation results (Boffetta and Sokolov \cite{Bof}); 
    \item[(b)] prevalence of power law scaling of RPD in 2D FDT enstrophy cascade (no matter how weak spatial intermittency effects are, as long as they are not negligible), corroborating the difficulty in observing Lin \cite{Lin} exponential scaling law in laboratory experiments (Jullien \cite{Jul});
  \end{itemize}
  \item[*] quasi-geostrophic FDT aspects exhibiting an enhanced RPD in the baroclinic regime of the energy cascade and a negative eddy-viscosity development to shed some insight into these aspects;
  \item[*] quasi-geostrophic FDT aspects exhibiting particle clumping in the baroclinic regime of the enstrophy cascade;
  \item[*] particle clustering due to compressibility effects in FDT, corroborating the laboratory experiment and numerical simulation results (Cressman et al. \cite{Cre}) where compressibility is effectively produced on the free surface of a shallow fluid layer.
\end{itemize}

\noindent These results are developed from the established scaling relations for the various physical FDT cases and are further validated via alternative dimensional/scaling developments for the various physical FDT cases similar to the one given for 3D FDT by Batchelor and Townsend \cite{Bat1}.

\vspace{0.3in}

\noindent\Large\textbf{2. 3D Relative Particle Dispersion}\\

\large Consider the temporal evolution of the separation between two particle trajectories in the inertial regime $\eta \ll R \ll L$ in high-$R_e$ FDT, $\eta$ being the Kolmogorov microscale. The equation for the separation distance between two particles is

\be\tag{5}
\frac{d}{dt} {\bf R} (t) = \textbf{V}(\textbf{R}, t)
\ee

\noindent from which,

\be\tag{6}
\frac{d}{dt} \lb {\bf R} (t) \rb^2 = 2 {\bf R} (t) \cdot \textbf{V}(\textbf{R}, t).
\ee

\noindent Assuming the scaling relation,

\be\tag{7}
\left| \textbf{V}(\textbf{R}, t) \right| \sim \lb R (t) \rb^\al, ~\al < 1
\ee

\noindent (symptomatic of a {\it non-Lipschitzian} behavior in the inertial range)\footnote{The non-differentiability of the velocity field indeed implies a stronger restriction on the {\it H\"{o}lder} scaling exponent $\alpha$: $\alpha \leq 1/3$ (see (11) below).} and averaging over many particles and many different initial separations, we obtain from (6),

\be\tag{8}
\frac{d}{dt} \la \lb R (t) \rb \ra^2 \sim \la \lb R (t) \rb^2 \ra^{\lp 1 + \al \rp/2}
\ee

\noindent from which,

\be\tag{9}
\la \lb R (t) \rb^2 \ra \sim t^{\frac{2}{1 - \al}}.
\ee

\noindent Observe that (9) shows a super-diffusive growth since $\al > - 1$ (for ordinary diffusion like the Brownian motion, $\al = -1$, so $\la \lb R (t) \rb^2 \ra \sim t$). For the K41 scaling given by $\al = 1/3$, (9) leads to the RO scaling,\footnote{For a chaotic system (with positive Liapunov exponent $\lambda$), we have

\be\notag
R (t) \sim e^{\lambda t}.
\ee

In FDT, there is a unique $V(R, t)$ for each $R$, so there is now a continuum of $\lambda (R)$. Consequently, $\la \lb R (t) \rb^2 \ra$ grows {\it algebraically} in FDT rather than exponentially.}

\be\tag{10}
\la \lb R (t) \rb^2 \ra \sim t^3.
\ee

\vspace{.3in}

\noindent\Large\textbf{3. 3D Relative Particle Dispersion: Effects of Spatial Intermittency}\\

\large The RO theory does not take into account the spatial intermittency in Lagrangian turbulence that was revealed by laboratory experiments (Mordant et al. \cite{mor})\footnote{Experimental evidence (Jullien et al. \cite{Jul2}) on RPD in 2D turbulence also indicated that this process is highly intermittent (see Section 4.1).} and numerical simulations (Yeung and Borgas \cite{yebo}). As is well known, spatial intermittency effects would cause systematic departures from the RO scaling law (10), which uses mean energy transfer rate. On the other hand, spatial intermittency effects are known to become more pronounced at small scales\footnote{The Lagrangian structure functions have been found (Mordant et al. \cite{mor}, Xu et al \cite{xu}) actually to show stronger intermittency corrections than their Eulerian counterparts.}, so the turbulence activity gets concentrated in smaller and smaller regions of space and the active region (called the dissipative structures) becomes strongly convoluted like a {\it fractal} (Mandelbrot \cite{Man})\footnote{In fact, Richardson \cite{Ric} had already pointed out the discontinuous nature of the velocity field in the atmosphere - it was (like the {\it Wierstrass} function) continuous but not differentiable.}. Indeed, the smaller the interparticle separation, the stronger the spatial intermittency effects become. The fractal aspects of FDT may be simulated in a first approximation by representing the dissipative structures via a homogeneous fractal with non-integer {\it Hausdorff} dimension $D_0$ (Frisch et al. \cite{Fri}). Using then, the result,

\be\tag{11}
\al = \frac{1}{3} + \frac{D_0 - 3}{3}
\ee

\noindent(9) leads to

\be\tag{12}
\la \lb R (t) \rb^2 \ra \sim t^{\left( \frac{6}{5-D_0} \right)} \sim t^{\lb 3 - 3 \lp \frac{3 - D_0}{5 - D_0} \rp \rb}.
\ee

\noindent (12) may also be deduced alternatively via a dimensional/scaling development \textit{\`a la} Batchelor and Townsend \cite{Bat1} (see Appendix A (ii)). Noting that $D_0 < 3$, (12) shows that the effect of spatial intermittency is to cause reduction in RPD, in agreement with the numerical results of Boffetta and Sokolov \cite{Bof}. On the other hand, (12) shows that the spatial intermittency effects, no matter how strong (i.e. even in the limit $D_0 \Rightarrow 0$), cannot change the super-diffusive nature of RPD, i.e.,

\be\tag{13}
\la \lb R (t) \rb^2 \ra \sim t^{2\nu}, ~\nu > 1/2
\ee

\noindent (see footnote 4). 

\vspace{.3in}

\noindent\Large\textbf{4. 2D Relative Particle Dispersion}\\

\large In 2D FDT, for pair separations larger than the energy injection scale but smaller than the integral scale (as in the inverse energy cascade), we have the scaling behavior,

\be\tag{14}
V (R) \sim R^{1/3}
\ee

\noindent which leads to the following growth relation,

\be\tag{15}
\frac{d}{dt} \la \lb R (t) \rb^2 \ra \sim \la \lb R (t) \rb^2 \ra^{2/3}
\ee

\noindent and hence, the mean square pair separation grows as (Kowaleski and Peskin \cite{Kow}),

\be\tag{10}
\la \lb R (t) \rb^2 \ra \sim t^3.
\ee

\noindent On the other hand, when the pair separation exceeds the integral scale $L$, the particles become uncorrelated and Brownian diffusion sets in with the following growth relation,

\be\tag{16}
\la \lb R (t) \rb^2 \ra \sim t.
\ee

\noindent Okubo \cite{Oku} analyzed oceanic experimental data and found the $R^{4/3}$ law for the turbulent relative diffusivity along with the $t^3$ law for the mean square pair separation, in agreement with (15) and (10), respectively. The laboratory experiments of Jullien et al. \cite{Jul2} also confirmed (10).

For pair separations much smaller than the energy-injection scale, the velocity field would be smooth. RPD in the 2D enstrophy cascade was considered by Lin \cite{Lin} who therefore gave for the turbulent relative diffusivity, the following scaling result,

\be\tag{17}
\frac{d}{dt} \la \lb R (t) \rb^2 \ra \sim \tau^{1/3} \la \lb R (t) \rb^2 \ra
\ee

\noindent and hence, for the mean square separation, the exponential growth behavior (controlled by the largest Liapunov exponent of the local smooth flow),

\be\tag{18}
\la \lb R (t) \rb^2 \ra \sim e^{\tau^{1/3} t}
\ee

\noindent where $\tau$ is the mean enstrophy dissipation rate. Implicit in (17) is the assumption that the effective shear acting across $R$ in the 2D enstrophy cascade arises from eddies of size much greater than $R$ and hence becomes spatially uniform (Kraichnan and Montgomery \cite{Kmont}). (17) may also be deduced alternatively via a dimensional/scaling development similar to the one given for 3D FDT by Batchelor and Townsend \cite{Bat1} (see Appendix A (iii)).

Balloon measurements in the atmosphere (Morel and Larcheveque \cite{Mor} and Er-El and Peskin \cite{Er-}) provided evidence comprising second-order structure functions and turbulent relative diffusivities, which were approximately proportional to the square of the separation length, and separation variances which grew approximately exponentially in time, hence supporting the existence of the enstrophy cascade and the Lin RPD scaling result (18) in 2D. Babiano et al. \cite{Bab} did numerical calculation of RPD in 2D FDT and found that, if the initial pair separation is larger than the energy-injection scale, RPD follows the $t^3$ law (10) up to the most energetic scales. On the other hand, RPD was found to follow the $e^{\tau^{1/3} t}$ law (18) only in a very short transient stage when the initial separation lies at the bottom of the enstrophy cascade. This was also confirmed by Kowaleski and Peskin \cite{Kow} via numerical calculations. Particle-path data from a float experiment (Ollitrault et al. \cite{Oll}) showed that the turbulent relative diffusivity varies as $R^2$ and $R^{4/3}$ for distances smaller and larger, respectively, than a forcing scale of the order of the Rossby radius of deformation. Two particles initially separated between 40 km and 300 km dispersed according to the $t^3$ law (10) (with turbulent relative diffusivity $\sim R^{4/3}$) while those with smaller initial separation distances dispersed according to the $e^{\tau^{1/3} t}$ law (18) (with turbulent relative diffusivity $\sim R^2$). The Lin scaling law (18) was confirmed at early times in a laboratory experiment by Jullien \cite{Jul}, but the range of pair separation scales over which this was observed was again very small, so this observation was rather difficult. This situation appears to be traceable to spatial intermittency effects in the enstropy cascade\footnote{Direct numerical simulations of freely decaying 2D FDT (McWilliams \cite{McW}, Benzi et al. \cite{Ben}, Brachet et al. \cite{Bra}, Kida \cite{Kid}, Ohkitani \cite{Ohk}, Schneider and Farge \cite{Sch}) and forced-dissipative 2D FDT (Basdeveant et al. \cite{Bas}, Legras et al. \cite{Leg}, Tsang et al. \cite{Tsa}) showed spatial intermittency in the enstrophy dissipation field caused by the presence of coherent structures.}, as shown in the Section 4.2 below. It is of interest to note that even the numerical investigations of Elhmaidi et al. \cite{Elh} and Zouari and Babiano \cite{Zou} showed that RPD in 2D energy cascade is affected by coherent structures and indeed exhibits steeper than the $t^3$ law, symptomatic of spatial intermittency corrections (see Section 4.1 below).

\vspace{.3in}

\noindent\textbf{4.1 Inverse Energy Cascade}\\

The spatial intermittency effects in the inverse energy cascade may again be incorporated via the fractal aspects of active turbulence regions. Thus, using the result (Frisch et al. \cite{Fri}),

\be\tag{19}
\al = \frac{3 - D_0}{3}
\ee

\noindent (9) leads to

\be\tag{20}
\la \lb R (t) \rb^2 \ra \sim t^{\left( \frac{6}{D_0} \right)}\sim t^{\lb 3 + \frac{3}{D_0} \lp 2 - D_0 \rp \rb}.
\ee

\noindent Noting that $D_0 < 2$, (20) shows that the effect of spatial intermittency in the energy cascade is to make RPD go steeper than the $t^3$ law, in agreement with the numerical calculations of Elhmaidi et al. \cite{Elh} and Zouari and Babiano \cite{Zou}\footnote{Indeed, RPD in the inverse energy cascade was found (Babiano et al. \cite{Bab}, Jullien et al. \cite{Jul2}), unlike Brownian motion, not to be a progressive process, but rather highly intermittent involving sequences of quiet periods and sudden bursts, so the two particles stay close together for a long time and then separate quite suddenly.}.

\vspace{.3in}

\pagebreak
\noindent\textbf{4.2 Enstrophy Cascade}\\

On incorporating the spatial intermittency effects in the enstrophy cascade as per the homogeneous fractal model for the enstrophy dissipation structures, we have (Shivamoggi \cite{Shi1}),

\be\tag{21}
\al = \frac{1 + D_0}{3}.
\ee

\noindent Using (21), (9) leads to

\be\tag{22}                              
\la \lb R (t) \rb^2 \ra \sim t^{\left( \frac{6}{2-D_0} \right)}\sim t^{\lb 3 + 3 \lp \frac{D_0}{2 - D_0} \rp \rb}.
\ee

\noindent The effect of spatial intermittency is to make RPD go slower than the Lin exponential growth law (18) which becomes operational in the space-filling limit $D_0 \Rightarrow 2$. On the other hand, this also shows that, in the presence of spatial intermittency, however small, the Lin exponential growth law (18) is replaced by the power law growth (22). This appears to clarify the difficulty in observing the Lin exponential growth regime in both laboratory experiments (Julien \cite{Jul}) and numerical simulations (Babiano et al. \cite{Bab}).

\vspace{.3in}

\noindent\Large\textbf{5. Quasi-geostrophic Relative Dispersion}\\

\large The dynamics of a 3D rapidly rotating fluid is characterized by the geostrophic balance between the Coriolis force and pressure gradient transverse to the axis of rotation. Quasi-geostrophic dynamics refers to the nonlinear dynamics governed by the first-order departure from this linear balance and is inherently 3D. The governing equation is the quasi-geostrophic potential vorticity equation (Charney \cite{Cha}) for an equivalent barotropic fluid in the $f$-plane ($f$ being the local Coriolis parameter). The term representing baroclinic effects in the flow in this equation introduces a characteristic length scale, namely, the Rossby radius of deformation $R_0 \equiv \sqrt{g H}/f$, into the problem ($H$ being the depth of the ocean taken to be uniform and $g$ being the acceleration due to gravity)\footnote{We are using the simplest mathematical model of large-scale, nearly horizontal oceanic motion incorporating the force of gravity and the Coriolis force due to the Earth's rotation, which is the one-layer homogeneous ocean with a uniform depth and a spherical free surface.}. Consequently, this problem exhibits some interesting departures from the properties of classical 2D turbulence (Shivamoggi \cite{Shi1}, \cite{Shi2}).

\vspace{.3in}

\noindent\textbf{5.1 Energy Cascade}\\

Upon incorporating the spatial intermittency effects in the energy cascade as per the homogeneous fractal model for the energy dissipation structures, we have (Shivamoggi \cite{Shi1})

\be\tag{23a, b}
\al = \lc
\begin{matrix}
\displaystyle \frac{4 - D_0}{3} ~, ~\ell_n \gg R_0\\
\\
\displaystyle \frac{3 - D_0}{3} ~, ~\ell_n \ll R_0.
\end{matrix}
\right.
\ee

\noindent Observe that (23b), which corresponds to the barotropic regime, is the same as the 2D hydrodynamics result (19).

Using (23), (9) leads to

\be\tag{24a, b}
\la \lb R (t) \rb^2 \ra \sim \lc
\begin{matrix}
t^{\lp \frac{6}{D_0 - 1} \rp} ~, ~\ell_n \gg R_0\\
t^{\lp \frac{6}{D_0} \rp} ~, ~\ell_n \ll R_0.
\end{matrix}
\right.
\ee

\noindent Observe that (24b) is again the same as the 2D hydrodynamic result (20). The effect of spatial intermittency in the energy cascade is, as in the 2D hydrodynamic case, to make RPD go steeper than the inertial-range scaling laws,

\be\tag{25a, b}
\la \lb R (t) \rb^2 \ra \sim \lc
\begin{matrix}
t^6 ~, ~\ell_n \gg R_0\\
t^3 ~, ~\ell_n \ll R_0.
\end{matrix}
\right.
\ee

\noindent Observe that RPD in the baroclinic regime $\lp \ell_n \gg R_0 \rp$, as per (25a), is greatly increased due to enhanced vortex stretching in this regime. This result may also be deduced alternatively via a dimensional/scaling development \textit{\`a la} Batchelor and Townsend \cite{Bat1} (see Appendix A (iv)).

Further insight into the unusual aspects of the quasi-geostrophic RPD problem in the inverse energy cascade may be obtained by considering the eddy viscosity development for quasi-geostrophic turbulence.

\vspace{.3in}

\noindent\textbf{5.2 Eddy Viscosity for the Inverse Energy Cascade}\\

Kraichnan \cite{Kra} proposed that the spontaneous development and net energy gain of large-scale structures in the inverse energy cascade from small-scale structures can be described by a negative eddy viscosity $\nu_T$ (which may be either introduced phenomenalogically or derived via closure approximations)\footnote{It should be mentioned that the concept of eddy viscosity is not on strong grounds because it is based on the idea that scales of motion of given size are acted on by smaller scales as if the latter were an augmentation of the equilibrium thermal agitation. This idea is not totally valid, thanks to the lack of clear separation between the two scale sizes.}. Following Kraichnan \cite{Kra}, one may treat the eddy viscosity as constant and calculate it by balancing the net eddy-viscous {\it gain} with the energy flux rate $\varepsilon$ {\it into} the explicit scales,

\be\tag{26}
\int\limits_0^{k_c} 2 k^2 \nu_T E (k) d k = -\varepsilon
\ee

\noindent $k_c$ being the cut-off wave number so the explicit scales are given by $k < k_c$. The negative sign on the right in (26) arises from the fact that energy is flowing toward smaller wavenumbers (rather than larger wavenumbers).

Using the energy spectra (Shivamoggi \cite{Shi1}),

\be\tag{27a, b}
E (k) \sim \lc
\begin{matrix}
c_k \eps^{2/3} k^{-5/3} ~, ~k R_0 \gg 1\\
c_k \eps^{2/3} R_0^{-2/3} k^{-7/3} ~, k R_0 \ll 1
\end{matrix}
\right.
\ee

\noindent $c_k$ being a constant, the energy balance relation (26) gives

\be\tag{28a, b}
\nu_T \sim \lc
\begin{matrix}
-\frac{2}{3} {c_k}^{-1} \eps^{1/3} {k_c}^{-4/3} ~, ~k_c R_0 \gg 1\\
-\frac{1}{3} {c_k}^{-1} \eps^{1/3} \lp k_c R_0 \rp^{2/3} {k_c}^{-4/3} ~, ~k_c R_0 \ll 1.
\end{matrix}
\right.
\ee

\noindent It is interesting to note that (27a, b) may be rewritten in the Leslie-Quarini \cite{Les} universal form,

\be\tag{29a, b}
\nu_T \sim \lc
\begin{matrix}
\displaystyle -\frac{2}{3} {c_k}^{-3/2} \sqrt{\frac{E \lp k_c \rp}{k_c}} ~, ~k_c R_0 \gg 1\\
\displaystyle -\frac{2}{3} \al {c_k}^{-3/2} \sqrt{\frac{E \lp k_c \rp}{k_c}} ~, ~k_c R_0 \ll 1.
\end{matrix}
\right.
\ee

\noindent (29a) corresponds to the negative eddy viscosity result for classical 2D turbulence given by Kraichnan \cite{Kra} while (29b) corresponds to the baroclinic regime - observe the explicit appearance, as one would expect, of the baroclinic parameter $\al$,

\be\tag{30}
\al \equiv k_c R_0
\ee

\noindent in the baroclinic regime\footnote{A similar result occurs for eddy viscosity in a compressible turbulence (Shivamoggi and Hussaini \cite{Shi3}) showing the explicit appearance of the Zakharov-Sagdeev \cite{Zak} compressibility parameter, $Z \equiv \rho c^3 / \varepsilon \ell$, $\rho$ being the mass density and $c$ being the speed of sound.}. (29) shows that the Leslie-Quarini universal form for the eddy viscosity has a certain robustness to it (as also indicated previously by Shivamoggi and Hussaini \cite{Shi3}). Observe further that the higher turbulent transport\footnote{This may be seen by rewriting (28) in the following form,

\be\notag
\nu_T \sim \lc
\begin{matrix}
-\frac{2}{3} {c_k}^{-1} \varepsilon^{1/3} {R_0}^{4/3} \al^{-4/3} ~, ~\al \gg 1\\
\\
-\frac{2}{3} {c_k}^{-1} \varepsilon^{1/3} {R_0}^{4/3} \al^{-2/3} ~, ~\al \ll 1.
\end{matrix}
\right.
\ee} in the baroclinic regime indicated by (28b) and (29b) is consistent with an enhanced RPD indicated by (25a).

For a discussion providing some insight into an actual physical mechanism underlying the negative eddy viscosity, see Appendix B.\\

\noindent\textbf{5.3 Enstrophy Cascade}\\

Upon incorporating the spatial intermittency effets in the enstrophy cascade as per the homogeneous fractal model for the enstrophy dissipation structures, we have (Shivamoggi \cite{Shi1})

\be\tag{31a, b}
\al = \lc
\begin{matrix}
\frac{2 + D_0}{3} ~, ~\ell_n \gg R_0\\
\\
\frac{1 + D_0}{3} ~, ~\ell_n \ll R_0.
\end{matrix}
\right.
\ee

\noindent Observe that (31b), which corresponds to the barotropic regime, is the same as the 2D hydrodynamics result (21).

Using (31), (9) leads to

\be\tag{32a, b}
\la \lb R (t) \rb^2 \ra \sim \lc
\begin{matrix}
t^{\lp \frac{6}{1 - D_0} \rp} ~, ~\ell_n \gg R_0\\
t^{\lp \frac{6}{2 - D_0} \rp} ~, ~\ell_n \ll R_0.
\end{matrix}
\right.
\ee

\noindent Observe that (32b) is the same as the 2D hydrodynamic result (22). The effect of spatial intermittency in the enstrophy cascade is again to make RPD go slower than the inertial-range scaling laws,

\be\tag{33a, b}
\la \lb R (t) \rb^2 \ra \sim \lc
\begin{matrix}
t^{-6} ~, ~\ell_n \gg R_0\\
e^{\tau^{1/3} t} ~, ~\ell_n \ll R_0.
\end{matrix}
\right.
\ee

\noindent (33a, b) may also be deduced alternatively via a dimensional/scaling development \textit{\`a la} Batchelor and Townsend \cite{Bat1} (see Appendix A (v)). The particle clumping  (implying turbulent {\it de-mixing}, see footnote 19) indicated in the baroclinic regime $\lp \ell_n \gg R_0 \rp$, as per (33a), may be understood by noting that the divorticity sheets are the enstrophy dissipation structures in the enstrophy cascade where the quasi-geostrophic turbulence activity is concentrated. The divorticity sheets are intensified by the enhanced vortex stretching in the baroclinic regime produced by the deformed free surface in the quasi-geostrophic dynamics, and on the other hand, are more likely to occur near vortex nulls (Shivamoggi et al. \cite{Shi4}, \cite{Shi5}) where particle clumping is favored to occur (see footnote 20). For the barotropic regime ($\ell_n \ll R_0$), Charney \cite{Cha2} showed that a potential enstrophy inertial range exists, in which RPD grows exponentially, as indicated by (33b).

\vspace{.3in}

\noindent\Large\textbf{6. Compressibility Effects on Relative Particle Dispersion}\\

\large Intuitively, fluid compressibility is believed to lead to trapping of particles for long times and counteracting their tendency to drift away from each other - strong fluid compressibility would lead to particle clustering (Falkovich et al. \cite{Fal}). Laboratory experiments and numerical simulations in full-fledged 3D compressible FDT are not at hand yet. Shallow fluid layer flows provide an interesting alternative in this regard because the horizontal divergence of the free-surface flow on a shallow fluid layer is non-zero even though the fluid is incompressible. Consequently, the motion of passive tracer particles\footnote{Particle tracers, which have the same density as that of the carrier fluid and very small size, can be approximated as point-like particles having the same velocity as that of the carrier fluid at the position of the particle. On the other hand, the effective velocity field for inertial particles can have a non-vanishing divergence even when they are advected by incompressible flows (Maxey and Riley \cite{Max}), Falkovich and Pumir \cite{fapu}). Consequently, inertial particles exhibit dissipative phase-space dynamics and hence, particle clustering, especially in strain-dominated regions.} (used as surface markers in oceans) is not representative of 3D incompressible dynamics because such particles will respond only to the fluid flow on the free surface and not to the flow normal to the surface. They sample the horizontal components of the flow-velocity field and hence exhibit a dissipative phase space dynamics involving asymptotic evolution on an attractor and provide a convenient framework to analyze the compressibility effects on RPD even though the flow velocity is very small compared with the speed of sound (Sommerer and Ott \cite{Som}). Laboratory experiments and numerical simulations have therefore investigated RPD on a free surface (Cressman et al. \cite{Cre}) to explore the compressibility effects on the RPD problem. On the other hand, Cressman et al. \cite{Cre} expressed the necessity to have a theoretical framework to explain their results. We now propose to provide one such theoretical formulation.

On assuming barotropic fluid and adiabatic flow processes, we obtain (Shivamoggi \cite{Shi6})

\be\tag{34}
\al = \frac{\gamma - 1}{3 \gamma - 1}
\ee

\noindent where the polytrope exponent $\gamma \lp 1 < \gamma < \infty \rp$ may be treated as a compressibility parameter (the limit $\gamma \Rightarrow \infty$ corresponds to the incompressible fluid and the limit $\gamma \Rightarrow 1$ corresponds to infinite compressibility).

Using (34), (9) leads to

\be\tag{35}
\la \lb R (t) \rb^2 \ra \sim t^{\lp 3 - 1/\gamma \rp}.
\ee

\noindent (35) may also be deduced alternatively via a dimensional/scaling development {\it \`a la} Batchelor and Townsend \cite{Bat1} (see Appendix A (vi)). (35) shows that the effect of compressibility is to cause a reduction in RPD (the power law growth now has a scaling exponent smaller than 3), in agreement with the laboratory experiments and numerical simulations (Cressman et al. \cite{Cre}). Physically, this may be traced to an enhanced effective shear\footnote{On using (A$\cdot$23b), observe that 

\be\notag
V/R \sim R^{- 2\gamma / (3\gamma - 1)} \sim R^{- 2/3 - 2/3(3\gamma - 1)} .
\ee} in compressible FDT (due to the tendency of vortices to become more resilient and stretch stronger in a compressible fluid (Shivamoggi \cite{Shi7})).

On the other hand, on incorporating the spatial intermittency effects in compressible turbulence (as indicated by the numerical simulations of Lele et al. \cite{Lee}, Passot et al. \cite{Pas}), and using a homogeneous fractal model for the kinetic energy dissipative structures, we have (Shivamoggi \cite{Shi8})

\be\tag{36}
\al = \lp \frac{\gamma - 1}{3 \gamma - 1} \rp \lp D_0 - 2 \rp.
\ee

\noindent Using (36), (9) leads to

\be\tag{37}
\la \lb R (t) \rb^2 \ra \sim t^{\lb \frac{3 - 1/\gamma}{1 + \lp \frac{\gamma - 1}{2 \gamma} \rp \lp 3 - D_0 \rp} \rb}.
\ee

\noindent Noting that $D_0 < 3$, (37) shows that the effect of spatial intermittency is to cause further reduction in RPD. On the other hand, on noting that the dissipative structures in a compressible turbulence are typically shock-wave like $\lp D_0 = 2 \rp$, (37) becomes

\be\tag{38}
\la \lb R (t) \rb^2 \ra \sim t^2 ~, \forall \gamma
\ee

\noindent indicating that RPD in intermittent compressible turbulence occurs in the ballistic regime. This appears to be physically plausible because, in the presence of shock waves, particle clustering\footnote{Thanks to a compressible-fluid flow like flow situation prevailing on the free-surface of shallow fluid layers, local convergence and divergence regions of particle density are observed there (Elhmaidi et al. \cite{Elh} and Cielsik et al. \cite{Cie}). In convergence regions, where particles tend to clump, there is downwelling and vice versa. Further, downwellings are found to occur near strain-dominated (thin elongated) regions while upwellings (associated with centrifugal action) are found to occur near rotation-dominated (patch like) regions.} renders the velocity increment become independent of the interparticle separation, and hence forcing RPD to go ballistic. It is of interest to note, in comparison with (38), that laboratory experiments (Cressman et al. \cite{Cre}) indicated a scaling exponent of 1.65 while numerical simulations (Cressman et al. \cite{Cre}) indicated a scaling exponent of 1.80. Further, (37) also shows that compressibility effects, no matter how strong, cannot change the super-diffusive nature of RPD. Indeed, in the infinite compressibility limit $\gamma \Rightarrow 1$, (37) yields

\be\tag{39}
\la \lb R (t) \rb^2 \ra \sim t^2 ~, ~\forall D_0
\ee

\noindent pertaining again to the ballistic regime (38)!

\vspace{0.3in}

\noindent\Large\textbf{7. Discussion}\\

\large In this paper, phenomenological developments are used to explain several aspects of RPD in different physical FDT situations. The role played by the FDT cascade physics underlying this process is investigated. Many of these aspects are motivated by previous laboratory experiment and numerical simulation results. These are,
\begin{itemize}
  \item[*] spatial intermittency effects exhibiting,
  \begin{itemize}
    \item[(a)] reduction of RPD in 3D FDT, corroborating the numerical simulation results (Boffetta and Sokolov \cite{Bof});
    \item[(b)] prevalence of power-law scaling of RPD in 2D FDT enstrophy cascade (no matter how weak spatial intermittency effects are, as long as they are not negligible), corroborating the difficulty in observing Lin \cite{Lin} scaling law in laboratory experiments (Jullien \cite{Jul}) and numerical simulations (Babiano et al. \cite{Bab} and Kowaleski and Peskin \cite{Kow});
  \end{itemize}
  \item[*] quasi-geostrophic FDT aspects exhibiting an enhanced RPD in the baroclinic regime of the energy cascade (some insight into this aspect has been attempted via a negative eddy-viscosity development);
  \item[*] quasi-geostrophic FDT aspects exhibiting particle clumping in the baroclinic regime of the enstrophy cascade (this aspect appears to be associated with the tendency of divorticity sheets to occur near the vortex nulls);
  \item[*] reduction of RPD, development of the ballistic regime and particle clustering due to compressibility effects in FDT, corroborating the laboratory experiment and numerical simulation results (Cressman et al. \cite{Cre}) where compressibility is effectively produced in the free-surface flow on a shallow fluid layer.
\end{itemize}

These results are direct consequences of the established scaling relations for the various physical FDT cases and are further validated via alternative dimensional/scaling developments similar to the one given for 3D FDT by Batchelor and Townsend \cite{Bat1}. \\

It should be mentioned, however, that the 3D RO scaling result (3) has received little experimental support due to the difficulty of performing Lagrangian measurements over a broad enough range of time and with sufficient accuracy. Even in recent laboratory experiments (Ott and Mann \cite{Ott}, Sawford \cite{Saw}, Bourgoin et al. \cite{Bour}, Salazar and Collins \cite{Sal}, Sawford and Pinton \cite{sapi}); with high-speed photography to track particles, and in numerical simulations (Yeung \cite{Yeu}) with the highest possible resolution possible for homogeneous isotropic turbulence, it is known to be hard to obtain an extended range with the RO scaling. The difficulty appears to be due to,
\begin{itemize}
  \item [*] shrinkage of the inertial range and enhancement of finite Reynolds number effects in the Lagrangian statistics (Sawford \cite{Saw1}, Mordant et al. \cite{mor1});
  \item[*] contamination of the inertial range by the usual dissipative effects at the ultraviolet end and by the external forcing effects at the infrared end of the spectrum caused by inadequate scale separation;
  \item[*] persistent memory of initial separation.
\end{itemize}

\vspace{.3in}

\noindent\Large\textbf{Acknowledgments}\\

\large Much of this work was carried out during the course of my participation in the Turbulence Workshop at the Kavli Institute for Theoretical Physics, Santa Barbara. I am thankful to Professors Katepalli Sreenivasan, Grisha Falkovich and Eberhard Bodenschatz for the invitation. I am also thankful to Professors Grisha Falkovich and Gert Jan van Heijst for valuable remarks. This research was supported in part by NSF grant No. PHY05-51164.

\vspace{.3in}

\noindent\Large\textbf{Appendix A. Dimensional/Scaling Developments}\\
\vspace{0.2in}
\noindent\Large \textbf{(i) 3D Relative Particle Dispersion}

\large Batchelor and Townsend \cite{Bat1} used dimensional arguments to postulate that the rate of change of the mean square interparticle separation in 3D FDT is given by

\be\tag{A$\cdot$1}
\frac{d}{dt} \la[R(t)]^2\ra \sim \varepsilon \tilde{t}^{\text{ 2}} \cdot f\left( \frac{R(t_0)}{\varepsilon^{1/2} \tilde{t}^{3/2}} , \  \frac{\varepsilon^{1/2} \tilde{t}}{\nu^{1/2}} \right)
\ee
where $\tilde{t} \equiv t - t_0$. For large $\tilde{t}$, the influence of the initial conditions (at $t=t_0$) vanishes, while in the large Reynolds number limit, the influence of fluid viscosity $\nu$ vanishes; (A$\cdot$1) then leads to
\be\tag{A$\cdot$2}
\frac{d}{dt} \la[R(t)]^2\ra \sim \varepsilon \tilde{t}^{\text{ 2}} \sim \varepsilon^{1/3} \la[R(t)]^2\ra^{2/3}
\ee 

\noindent in agreement with (4).

\vspace{.3in}

\noindent\Large \textbf{(ii) Spatially Intermittent 3D Relative Particle Dispersion} \\

\large On assuming that the dissipative structures may be represented by a homogeneous fractal with a non-integer Hausdorff dimension $D_0$, the energy dissipation rate is given by (Frisch et al. \cite{Fri}),

\be\tag{A$\cdot$3}
\varepsilon \sim \left( \frac{V^3}{R} \right) R^{3 - D_0} .
\ee

\indent Noting that

\be\tag{A$\cdot$4}
\tilde{t} \sim \frac{R}{V}
\ee

\noindent and using (A$\cdot$3), the rate of change of the mean square interparticle separation, as per dimensional arguments, is now given by

\be\tag{A$\cdot$5}
\frac{d}{dt} \la[R(t)]^2\ra \sim \varepsilon^{\left( \frac{2}{5-D_0} \right)} \tilde{t}^{\left( \frac{1+D_0}{5-D_0} \right)} \cdot f\left( \frac{R(t_0)}{\varepsilon^{1/2} \tilde{t}^{3/2}}, \ \frac{\varepsilon^{1/2} \tilde{t}}{\nu^{1/2}} \right) .
\ee

\noindent In the large $\tilde{t}$ and the large Reynolds number limit, (A$\cdot$5) leads to

\be\tag{A$\cdot$6}
\frac{d}{dt} \la[R(t)]^2\ra \sim \varepsilon^{\left( \frac{2}{5-D_0} \right)} \tilde{t}^{\left( \frac{1+D_0}{5-D_0} \right)} \sim \varepsilon^{1/3} \la[R(t)]^2\ra^{\left( \frac{1+D_0}{6} \right)}
\ee

\noindent which, in the space-filling limit, $D_0 \to 3$, reduces to (A$\cdot$2). (A$\cdot$6) leads to 

\be\tag{A$\cdot$7}
\la[R(t)]^2\ra \sim \varepsilon^{\left( \frac{2}{5-D_0} \right)} \tilde{t}^{\left( \frac{6}{5-D_0} \right)}
\ee

\noindent in agreement with (12).

\vspace{.3in}

\noindent\Large\textbf{(iii) 2D Relative Particle Dispersion: Enstrophy Cascade} \\

\large The rate of change of the mean square interparticle separation, as per dimensional arguments, is now given by

\be\tag{A$\cdot$8}
\frac{d}{dt} \la[R(t)]^2\ra \sim \la[R(t)]^2\ra \tau \tilde{t}^{\text{ 2}} \cdot f\left( \frac{R(t_0)}{\nu^{1/2} \tilde{t}^{1/2}}, \ \tau^{2/3} \tilde{t}^{\text{ 2}} \right) .
\ee

\noindent In the large Reynolds number limit, on taking

\be\tag{A$\cdot$9}
\lim_{\nu \to 0} f\left( \frac{R(t_0)}{\nu^{1/2} \tilde{t}^{1/2}}, \ \tau^{2/3} \tilde{t}^{\text{ 2}} \right) \sim \frac{1}{\tau^{2/3} \tilde{t}^{\text{ 2}}}
\ee

\noindent (so as to make the effective shear acting across $R$ become spatially uniform), (A$\cdot$8) leads to

\be\tag{A$\cdot$10}
\frac{d}{dt} \la[R(t)]^2\ra \sim \tau^{1/3} \la[R(t)]^2\ra 
\ee

\noindent in agreement with (17).

\vspace{.3in}

\noindent\Large \textbf{(iv) Quasi-geostrophic Relative Particle Dispersion: Energy \\
Cascade} \\

\large The rate of change of the mean square interparticle separation, as per dimensional arguments, is now given by

\be\tag{A$\cdot$11}
\frac{d}{dt} \la[R(t)]^2\ra \sim \varepsilon \tilde{t}^{\text{ 2}} \cdot f\left( \frac{R(t_0)}{\varepsilon^{1/2} \tilde{t}^{3/2}}, \ \frac{R_0}{\varepsilon^{1/2} \tilde{t}^{3/2}}, \ \frac{\varepsilon^{1/2} \tilde{t}}{\nu^{1/2}} \right) .
\ee

\noindent In the large $\tilde{t}$, the large Reynolds number and the baroclinic limit, on taking 

\be\tag{A$\cdot$12}
\lim_{\tilde{t} \to \infty} \lim_{\nu \to 0} f\left( \frac{R(t_0)}{\varepsilon^{1/2} \tilde{t}^{3/2}}, \ \frac{R_0}{\varepsilon^{1/2} \tilde{t}^{3/2}}, \ \frac{\varepsilon^{1/2} \tilde{t}}{\nu^{1/2}} \right) \sim \frac{1}{(R_0 / \varepsilon^{1/2} \tilde{t}^{3/2})^2}
\ee

\noindent (A$\cdot$11) leads to

\be\tag{A$\cdot$13}
\frac{d}{dt} \la[R(t)]^2\ra \sim \varepsilon^2 R_0 ^{-2} \tilde{t}^{\text{ 5}} \sim \varepsilon^{1/3} R_0 ^{-1/3} \la[R(t)]^2\ra ^{5/6}
\ee

\noindent from which,

\be\tag{A$\cdot$14}
\la[R(t)]^2\ra \sim \varepsilon^2 R_0^{-2} \tilde{t}^{\text{ 6}}
\ee

\noindent in agreement with (25a). On the other hand, comparison of (A$\cdot$13) with (6) leads to

\be\tag{A$\cdot$15}
V(R) \sim \varepsilon^{1/3} R_0^{-1/3} \la R^2 \ra^{1/3}
\ee

\noindent in agreement with (27b).

\vspace{.15in}

\noindent\Large \textbf{(v) Quasi-geostrophic Relative Particle Dispersion: Enstrophy\\
Cascade} \\

\large Noting that the potential enstrophy for quasi-geostrophic flows in the Charney \cite{Cha} model is given by

\be\tag{A$\cdot$16}
U \sim \frac{V^2}{R^2} \sim \frac{\phi^2}{R^2} \left( \frac{1}{R^2} + \frac{1}{R_0 ^2} \right) 
\ee

\noindent the enstrophy dissipation rate is given by

\be\tag{A$\cdot$17a, b}
\tau \sim \left\{ \begin{matrix}
V^3 / R^3 ,& R \ll R_0 \\
V^3 R_0 / R^4 ,& R \gg R_0
\end{matrix} \right.
\ee

\noindent $\phi$ being the stream function.

Using (A$\cdot$4), (A$\cdot$17a, b) leads to

\be\tag{A$\cdot$18a, b}
\tau \sim \left\{ \begin{matrix}
\tilde{t}^{-3} ,& R \ll R_0 \\
\left( R_0 / R \right) \tilde{t}^{-3} ,& R \gg R_0 .
\end{matrix} \right.
\ee

\noindent The rate of change of the mean square interparticle separation, as per dimensional arguments, is given by

\be\tag{A$\cdot$19}
\frac{d}{dt} \la[R(t)]^2\ra \sim \la[R(t)]^2\ra \tilde{t}^{-1} \cdot f\left( \frac{R(t_0)}{\nu^{1/2} \tilde{t}^{1/2}}. \ \frac{R_0}{\nu^{1/2} \tilde{t}^{1/2}} \right)
\ee

\indent For the barotropic regime ($R \ll R_0$), using (A$\cdot$18a), one obtains the previous result (A$\cdot$8). On the other hand, for the baroclinic regime ($R \gg R_0$), using (A$\cdot$18b), (A$\cdot$19) becomes

\be\tag{A$\cdot$20}
\frac{d}{dt} \la[R(t)]^2\ra \sim R_0 ^2 \tau^{-2} \tilde{t}^{-7} \cdot f\left( \frac{R(t_0)}{\nu^{1/2} \tilde{t}^{1/2}}, \ \frac{R_0}{\nu^{1/2} \tilde{t}^{1/2}} \right) .
\ee

\noindent In the large Reynolds number limit, (A$\cdot$20) leads to

\be\tag{A$\cdot$21}
\frac{d}{dt} \la[R(t)]^2\ra \sim \left( \frac{R_0}{\tau}\right)^2 \tilde{t}^{-7} \sim \left( \frac{R_0}{\tau}\right)^{-1/3} \la[R(t)]^2\ra^{-7/6}
\ee

\noindent from which,

\be\tag{A$\cdot$22}
\la[R(t)]^2\ra \sim \left( \frac{R_0}{\tau} \right)^2 \tilde{t}^{-6} , \ R \gg R_0
\ee

\noindent in agreement with (33a).

\vspace{.3in}

\noindent\Large \textbf{(vi) Compressible Relative Particle Dispersion} \\

\large On assuming barotropic fluid and adiabatic flow processes, we have the following scaling relations (Shivamoggi \cite{Shi6}),

\be\tag{A$\cdot$23a, b}
\left. \begin{matrix}
\rho(R) \sim R^{\left( \frac{2}{3\gamma - 1} \right)} \\
V(R) \sim R^{\left( \frac{\gamma - 1}{3\gamma - 1} \right)} .
\end{matrix} \right\}
\ee

\noindent $\rho$ being the mass density of the fluid and $\gamma$ being the polytrope exponent ($1 < \gamma < \infty$).

Using (A$\cdot$4), (A$\cdot$23a, b) leads to

\be\tag{A$\cdot$24}
\rho(\hat{t}) \sim \tilde{t}^{\text{ $1/\gamma$}} .
\ee

\noindent The rate of change of the mean square interparticle separation, as per dimensional arguments, is now given by

\be\tag{A$\cdot$25}
\frac{d}{dt} \la[R(t)]^2\ra \sim \frac{\hat{\varepsilon}}{\rho} \tilde{t}^{\text{ 2}} \cdot f\left( \frac{R(t_0) \rho^{1/2}}{\hat{\varepsilon}^{1/2} \tilde{t}^{3/2}} , \ \frac{ \hat{\varepsilon}^{1/2} \tilde{t} }{ \rho^{1/2} \nu^{1/2} }\right)
\ee

\noindent $\hat{\varepsilon}$ being the mean kinetic energy dissipation rate. In the large $\tilde{t}$ and the large Reynolds number limit, on using (A$\cdot$24), (A$\cdot$25) leads to

\be\tag{A$\cdot$26}
\frac{d}{dt} \la[R(t)]^2\ra \sim \hat{\varepsilon} \tilde{t}^{\left( 2 - 1/\gamma \right)} \sim \hat{\varepsilon}^{\left( \frac{\gamma}{3\gamma-1} \right)} \la[R(t)]^2\ra^{\left( \frac{2\gamma -1}{3\gamma -1} \right)}
\ee

\noindent which reduces to (A$\cdot$2) in the incompressible limit ($\gamma \to \infty$). (A$\cdot$26) leads to

\be\tag{A$\cdot$27}
\la[R(t)]^2\ra \sim \hat{\varepsilon} \tilde{t}^{\left( 3 - 1/\gamma \right)}
\ee

\noindent in agreement with (35).

\vspace{.3in}

\noindent\Large\textbf{Appendix B. Phenomenological Derivation of Negative Eddy Viscosity}\\

\large A sound phenomenological demonstration of negative eddy viscosity via a simple model that captures the essential physics is apparently not at hand yet despite Kraichnan's \cite{Kra} spirited attempts motivated by complicated analytical developments. Analytical approaches used by Sivashinsky et al. \cite{Siv}, \cite{Siv2} and Gama et al. \cite{Gam} are restricted to flows possessing special symmetries. We propose to contribute toward this quest and give a simple phenomenological derivation of negative eddy viscosity for 2D energy cascade.

We borrow from a basic idea underlying the physical mechanism producing negative eddy viscosity proposed by Kraichnan \cite{Kra} - the large scales strain the small scales while a secondary flow associated with small scales grows. Further, the deterministic large-scale flow and a random small-scale vorticity field are characterized by disparate scales (Tur et al. \cite{Tur} and Fidutenko \cite{Fid}). Consider therefore a random homogeneous small-scale vorticity field $\boldsymbol{\omega}$ superposed on a stationary large-scale flow with stream function $\Psi$ and the concomitant velocity given by

\be\tag{B$\cdot$1}
{\bf V} = {\ihat}_z \times \na \Psi.
\ee

\noindent Linearizing about the large-scale flow (in analogy with a phenomenological development for MHD turbulence, sketched by Pouquet \cite{Pou}) we obtain, for the small-scale flow, with velocity {\bf v},

\be\tag{B$\cdot$2}
\frac{\pa {\bf v}}{\pa t} = \omega \na \Psi
\ee

\noindent where,

\be\notag
{\bs\omega} \equiv \na \times {\bf v} = \omega {\ihat}_z.
\ee

Assuming a normal-mode type evolution in time for the small-scale flow,

\be\tag{B$\cdot$3}
{\bf v} \sim e^{\sigma t}
\ee

\noindent so $1/\sigma$ may be interpreted as a coherence time of the small-scale flow, equation (B$\cdot$2) becomes

\be\tag{B$\cdot$4}
{\bf v} = \frac{\omega}{\sigma} \na \Psi.
\ee

On the other hand, the mean value of the large-scale vorticity $\Omega$ evolves according to

\be\tag{B$\cdot$5}
\frac{\pa}{\pa t} \la \Omega \ra + \na \cdot \la {\bf v} \omega \ra = 0.
\ee

Assuming the large-scale flow to comply with the Beltrami condition\footnote{A similar assumption was also made in the negative-viscosity development of Sivashinsky et al. \cite{Siv}, \cite{Siv2}.},

\be\tag{B$\cdot$6}
\la \Omega \ra = \la \Psi \ra/L^2
\ee

\noindent $L$ being a constant (of dimension length), and using equation (B$\cdot$4), equation (B$\cdot$5) becomes,

\be\tag{B$\cdot$7}
\frac{\pa}{\pa t} \la \Psi \ra = -\frac{L^2}{\sigma} \la \omega^2 \ra \na^2 \Psi
\ee

\noindent which implies that the nonlinear evolution of the large-scale flow is characterized by a negative eddy viscosity given by,

\be\tag{B$\cdot$8}
\nu_T \sim -\frac{L^2}{\sigma} \la \omega^2 \ra.
\ee

On the other hand, on recognizing that turbulent transport in the 2D inverse cascade is actually a competition between flow advection and vortex coalescence, (B$\cdot$8) may be rewritten as

\be\tag{B$\cdot$9}
\nu_T \sim \frac{1}{\sigma} \lp \la u^2 \ra - L^2 \la \omega^2 \ra \rp.
\ee

A phenomenological derivation of the flow advection term in (B$\cdot$9) may be given as follows. Consider a flow velocity ${\bf v} = \la u, v \ra$ with a displacement vector ${\bs\ell} = \la \ell_1, \ell_2 \ra$. We have, on complying with the continuity equation ({\it \`{a} la} Batchelor \cite{Bat} and Kraichnan \cite{Kra2}),

\be\tag{B$\cdot$10a, b}
\left.
\begin{matrix}
u \sim d \ell_1/dt \sim \sigma \ell_1\\
v \sim d \ell_2/dt \sim -\sigma \ell_2
\end{matrix}
\rc
\ee

\noindent while the vorticity, on using (B.10a,b), is given by

\be\tag{B$\cdot$11}
\omega \sim \frac{v}{\ell_1} - \frac{u}{\ell_2} \sim -u \ell_2 \lp \frac{1}{\ell_1^2} + \frac{1}{\ell_2^2} \rp.
\ee

The eddy viscosity (or turbulent relative diffusivity), on using (B.10a,b), is given by

\be\tag{B$\cdot$12}
\nu_T \sim \frac{1}{2} \frac{d}{dt} \la \ell^2 \ra \sim \sigma \lp \la \ell_1^2 \ra - \la \ell_2^2 \ra \rp.
\ee

\noindent Using (B$\cdot$10) and (B$\cdot$11), and prescribing,

\be\tag{B$\cdot$13}
L^2 \equiv \frac{\la \ell_1^2 \ra}{\lp 1 + \la \ell_1^2 \ra/\la \ell_2^2 \ra \rp^2}.
\ee

\noindent (B$\cdot$12) may be rewritten as, 

\be\tag{B$\cdot$14}
\nu_T \sim \frac{1}{\sigma} \lp \la u^2 \ra - L^2 \la \omega^2 \ra \rp
\ee

\noindent in agreement with (B$\cdot$9).

Observe that, according to (B$\cdot$14), the negative eddy viscosity effect disappears if the small-scale flow decorrelates rapidly in time (i.e., $\sigma \Rightarrow \infty$), as Kraichnan \cite{Kra} pointed out.

\end{document}